# Dynamic Force Spectroscopy: Looking at the Total Harmonic Distortion

Robert W. Stark[1]

*Center for Nanoscience and Ludwig-Maximilians-Universität München, Section Crystallography, Theresienstr. 41, 80333 München, Germany*

**Abstract.** Tapping mode atomic force microscopy is a standard technique for inspection and analysis at the nanometer scale. The understanding of the non-linear dynamics of the system due to the tip sample interaction is an important prerequisite for a correct interpretation data acquired by dynamic AFM. Here, the system response in tapping-mode atomic force microscope (AFM) simulated numerically. In the computer model the AFM microcantilever is treated as a distributed parameter system. With this multiple-degree-of-freedom (MDOF) approach the the total harmonic distortion in dynamic AFM spectroscopy is simulated.

## INTRODUCTION

The atomic-force microscopy (AFM) has become an standard inspection and analysis tool in research as well as in industry. The tapping or intermittent contact mode is presently the most widely used imaging modes in practical AFM applications. In this mode of operation, the forced oscillation amplitude of the force sensor is adjusted to a value between 10 nm and 100 nm. During imaging the amplitude is limited by the specimen surface which can be understood as a non-linear mechanical controller limiting the amplitude. Thus, a theoretical description of the system dynamics in this mode requires an understanding of the non-linear system dynamics [1]. The non-linear tip-sample interaction leads to a complicated system behavior. It was shown that the system is well behaved for a large set of parameters but that it also can exhibit a complex dynamics [2-4]. The non-linearity also induces higher harmonics in the system response which are amplified by the higher eigenmodes of the force sensor [5-8].

These higher harmonics can be measured by dynamic force spectroscopy recording the full spectral response of the system. This also allows one to directly measure transient tip-sample interaction forces by signal inversion [9]. In a simplified analysis the individual higher harmonics characterize the system dynamics [10].

The total harmonic distortion is a measure for the degree of the generation of higher harmonics. In the following, the response of the total harmonic distortion to the variation of average tip-sample gap is investigated by numerical simulations. To model the higher eigenmodes of the cantilever a 6-th-order state space model is used.

[1] E-mail: stark@nanomanipulation.de. This work was supported by the German Federal Ministry of Education and Research (BMBF) under Grant 03N8706

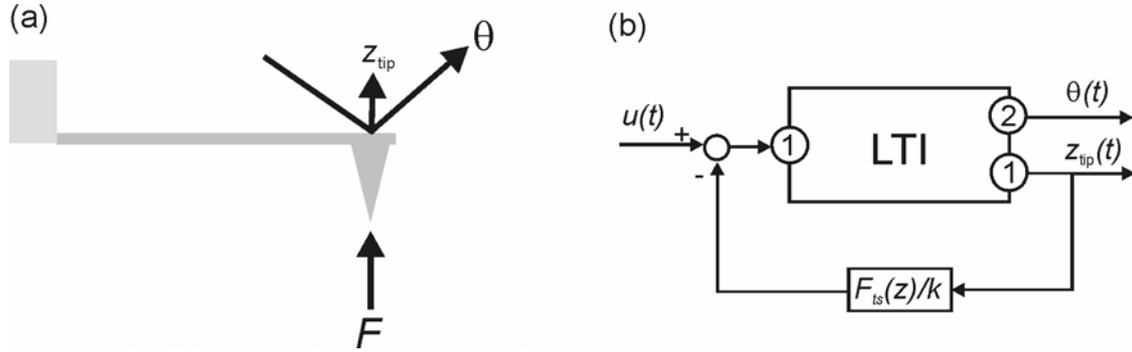

**FIGURE 1.** (a) Scheme of an atomic force microscope and (b) its representation by a linear system with a non-linear output feedback. The force F acting onto the tip is fed into the linear time invariant system. The tip position and the light lever readout are the system outputs.

# MODELLING

The microcantilever in atomic-force microscopy is approximated by a linear and time invariant (LTI) system. In tapping mode, its deflection is typically of the order of a few nanometers, whereas its thickness is in the range of microns. Therefore, the cantilever [Fig. 1 (a)] is modelled as a LTI system with a non-linear output feedback [Fig. 1 (b)]. To investigate the basic phenomena in the following, only forces that act onto the tip at input (1) will be considered..

Approximating the microcantilever by a $n$ degrees-of-freedom ($n$ eigenmodes) LTI system the equations of motion are given in state-space form by

$$\dot{\mathbf{x}} = \mathbf{A}\mathbf{x} + \mathbf{B}u, \tag{1}$$

$$\mathbf{y} = \mathbf{C}\mathbf{x}. \tag{2}$$

The time dependent state-vector $\mathbf{x} = (x_1, \dot{x}_1, \cdots, x_n, \dot{x}_n)$ contains the modal displacements and velocities. The system matrix

$$\mathbf{A} = \begin{bmatrix} \Phi_1 & 0 & 0 \\ 0 & \ddots & 0 \\ 0 & 0 & \Phi_n \end{bmatrix}, \text{ with } \Phi_i = \begin{bmatrix} 0 & 1 \\ -\hat{\omega}_i^2 & -2\gamma_i\hat{\omega}_i \end{bmatrix} \tag{3}$$

is a $2n \times 2n$ matrix. It consists of $2 \times 2$ submatrices $\Phi_i$ along the diagonal. The matrices $\Phi_i$ characterize the individual eigenmodes of the weakly damped system. The eigenfrequencies $\hat{\omega}_i = \omega_i / \omega_1$ are normalized to the fundamental resonance frequency, the modal damping is $\gamma_i$. In the case of heavy damping as it is the case for example in a liquid environment matrix $A$ also contains non-diagonal elements. The input vector is

$$\mathbf{B} = \begin{bmatrix} 0, \varphi_1(\xi_{tip})/M_1, \cdots, 0, \varphi_n(\xi_{tip})/M_n \end{bmatrix}^T. \tag{4}$$

It contains the modal deflection at the tip $\varphi_i(\xi_{tip})$ which is normalized by the generalized modal mass $M_i = \int_0^1 m\varphi_i(\xi)^2 d\xi$. Scalar $u$ is the input to the model, i.e. the driving force minus the tip-sample interaction force.

The components of the output vector $\mathbf{y} = [y_1, y_2]^T$, i.e. the tip displacement output $y_1$ that is used for feedback and the photodiode signal output $y_2$, are linear combinations of the states as defined in the output matrix

$$\mathbf{C} = \begin{bmatrix} \varphi_1(\xi_{tip})/n_{pos} & 0 & \cdots & \varphi_n(\xi_{tip})/n_{pos} & 0 \\ \varphi_1'(\xi_{sens})/n_{sig} & 0 & \cdots & \varphi_n'(\xi_{sens})/n_{sig} & 0 \end{bmatrix}. \quad (5)$$

The tip deflection output (1) is normalized with $n_{pos}$ to obtain a unit DC gain, i.e. it is normalized to a quasi-static spring constant $\hat{k}_{cant} = 1$ of the system at $\omega = 0$. The optical lever sensor output (2) is normalized by $n_{sig}$ to a unit response at $\omega = 0$.

The tip displacement $y_1 = \sum_{i=1}^{n} x_{2i-1}$ at output (1) is used to calculate the non-linear tip-sample interaction force $F_{ts}(y_1 - z_s)/k$. The resulting force is fed back to input (1) of the model.

The attractive part of the interaction force $(y_1 - z_s \geq a_0)$ is modelled as a van der Waals interaction force. A Derjaguin-Müller-Toporov (DMT) model [11] was used in the repulsive regime $(y_1 - z_s < a_0)$. Thus, the interaction force is

$$F_{ts}(y_1) = \begin{cases} -HR/\left[6(y_1 - z_s)^2\right] & y_1 - z_s \geq a_0 \\ -HR/6a_0^2 + \frac{4}{3}E^*\sqrt{R}(a_0 - y_1 + z_s)^{3/2} & y_1 - z_s < a_0 \end{cases}, \quad (6)$$

where $H$ is the Hamaker constant, $R$ the tip radius, and $a_0$ an interatomic distance. The effective contact stiffness is given by $E^* = \left[(1-\nu_t^2)/E_t + (1-\nu_s^2)/E_s\right]^{-1}$, where $E_t$ and $E_s$ are the respective elastic moduli and $\nu_t$ and $\nu_s$ the Poisson ratios of tip and sample.

As numerical parameters typical values for a beam shaped cantilever were used. Three eigenmodes were considered $(n = 3)$ for the computation of the system response. The modal deflection $\varphi_n$ and deflection angle $\varphi_n'$ were calculated from the well known eigenmodes of a uniform beam [12]. The tip and laser spot were assumed to be collocated at the end of the cantilever beam $\xi_{tip} = \xi_{sens} = 1$. The damping was set to $\gamma_i = 0.0025$ for all modes. Further parameters were: $k = 10 \ Nm^{-1}$, $R = 15 \ nm$, $E_t = 129 GPa$, $\nu_t = 0.28$, $E_s = 70 GPa$, $\nu_s = 0.3$, $a_0 = 0.166 nm$, and $H = 6.4 \times 10^{-20} J$. The driving frequency was $\omega = 1.0$, the amplitude of the driving force was $F_{dr} = 0.97 \ nN$, resulting in a free amplitude of $A_0 = 20 \ nm$. The simulation was implemented in MATLAB RELEASE 13 using SIMULINK (The Mathworks Inc., Natick, MA, USA).

# TOTAL HARMONIC DISTORTION

In order to compute the system response in a dynamic AFM spectroscopy experiment the sample position $z_s$ was reduced by ramping. At each approach step the ramp was halted and the system was allowed to equilibrate for more than 1000 cycles before data was extracted for Fourier transform (FFT) analysis. For $z_s \leq -3nm$ data was extracted every 0.5 nm, for larger $z_s$ the distance was decreased to 0.2 nm to capture the complex dynamics at small distances. Figure 2 shows the evolution of the amplitude and phase of the first harmonic (fundamental) together with the total harmonic distortion of the position output (1) and the average force. The harmonic distortion is defined by $THD = \left(\sum_{n=2}^{\infty} |c_n|^2\right)^{-1/2} / \left(\sum_{n=1}^{\infty} |c_n|^2\right)^{-1/2}$, where $|c_n|$ is the FFT amplitude of the $n$-th harmonic. It gives the fraction of power that is transferred into the higher harmonics as compared to the total power.

Far away from the sample, the oscillation amplitude of the fundamental is $2|c_1| = -20nm$, the phase is at $-90°$. There is only a very small average attractive force and a very small total harmonic distortion. Approaching to $z_s = -20nm$ the system is in the net attractive (low amplitude) regime as can be seen by the net-negative interaction force. With increasing strength of the attractive interaction the THD of the output signal also increases. Between $z_s = -18.5nm$ and $z_s = -18nm$ the system transits to the high amplitude state (arrows). This transition prevails in the phase as well as in the average interaction force. It is also visible in the THD which increases by 50%. Approaching further, the dynamics of the system changes at $z_s = -2.8nm$. The THD decreases significantly and recovers at $z_s = -1nm$ before it drops to zero. This behavior can be explained by the generation of subharmonics where spectral power is transferred into subharmonics.

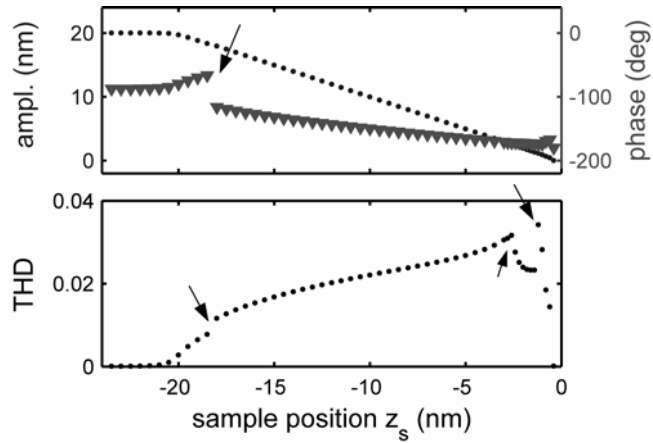

**Figure 2.** (Above) Amplitude and phase of the first harmonic. The transition from the low amplitude state into the high amplitude state can be identified by the phase jump (arrow). (Below) Total harmonic distortion of the position output. The transition between both states is accompanied by a step in the THD (arrow, left). At a small $z_s$ subharmonics are generated, which first leads to a reduction of the THD followed by a final maximum (arrows, right).

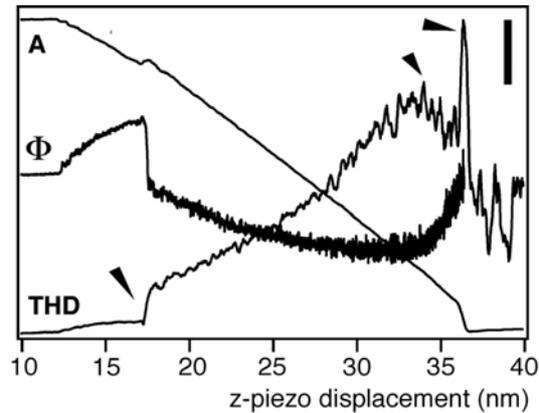

**Figure 3**. Experimental approach curves on a silicon sample. Amplitude $A$, phase $\Phi$ of the fundamental, and the total harmonic distortion $THD$. The arrows indicate the transition from the attractive to the repulsive branch (left) and both maxima in the THD (right). Scale bar: $A$: a.u., $\Phi$: 36°, $THD$: 4 %. (From Ref. [10]. (c) 2000 AIP, reprinted with permission).

## CONCLUSIONS

In comparison with experimental data obtained earlier [10] the characteristics of the response of the simulated $THD$ are similar to that of the experimental data in Fig. 3 although the scaling is different. The increase in the $THD$ at the transition from the attractive to the repulsive state (arrow) was also observed in the experimental data. Additionally, both maxima in Fig. 3 of the $THD$ (arrows) are well reproduced by the numerical simulations. This indicates that the numerical simulations capture basic features of the dynamics in tapping mode AFM. In order to achieve a better match of the numerical simulations to the experimental data a more precise modelling is necessary. This includes a better mathematical model for the cantilever as it can be obtained e.g. by system identification and an more precise model of the contact mechanics.